%% file: main.tex
\documentclass[prl,twocolumn,showpacs,superscriptaddress,10pt,noeprint]{revtex4-1} 
\usepackage[T1]{fontenc}
\usepackage[utf8]{inputenc}
\usepackage{bm,dcolumn}
\usepackage{amsmath,amsfonts,amssymb,amsthm,amscd} 
\usepackage{graphicx}
\usepackage{verbatim}
\usepackage[section]{placeins}
\usepackage{color}
\usepackage{soul}
\usepackage{xparse}
\usepackage{physics}
\usepackage{siunitx}
\usepackage{xcolor}
\usepackage{xspace}
\usepackage{lipsum}

\newcommand{\cmnt}[1]{\footnote{[Cmnt: {\footnotesize \textcolor{gray}{ #1} } ]}}
\newcommand{\cmntc}[1]{[Cmnt: \textcolor{brown}{ #1} ]}

\newcommand{\todoc}[1]{[\textcolor{gray}{\textbf{To do:}}   \textcolor{gray}{#1}]}
\newcommand{\figref}[1]{Fig.~\ref{#1}}
 
\renewcommand{\todoc}[1]{}    \renewcommand{\cmnt}[1]{} 		 \renewcommand{\cmntc}[1]{} 
\newcommand{\rs}{\rm \scriptscriptstyle}
\newcommand{\remove}[1]{}
							
\def\MHz{\text{MHz}\xspace}

\def\Om{\Omega\xspace}

\def\paper{Letter\xspace}

\newcommand\hbarOne{}
\newcommand\hbarSq{}
\newcommand\subwavelength{subwavelength\xspace}
\newcommand\lightshift{light-shift\xspace}
\newcommand\lightshifts{light-shifts\xspace}

\def\Rf{blue-detuned AC-Stark\xspace} 
\def\RfCapital{Blue-detuned AC-Stark\xspace} 
\def\Yavuz{red-detuned AC-Stark\xspace} 
\def\YavuzCapital{Red-detuned AC-Stark\xspace}
\def\EIT{EIT\xspace}

\def\RfShort{\text{blue-ac}}
\def\YavuzShort{\text{red-ac}}

\def\EITShort{\EIT}

\def\UNA{ U\xspace} \def\nn{\nonumber}
\def\ER{E_\sigma\xspace}

\def\beqa{\begin{eqnarray}}
\def\eeqa{\end{eqnarray}}

\newcommand{\beginsupplement}{        \setcounter{table}{0}
        \renewcommand{\thetable}{S\arabic{table}}        \setcounter{figure}{0}
        \renewcommand{\thefigure}{S\arabic{figure}}     }
     
\begin{document}
\title{Coherent optical nano-tweezers for ultra-cold atoms}

\author{P. Bienias}
\email{bienias@umd.edu}
\affiliation{Joint Quantum Institute, NIST/University of Maryland, College Park, Maryland 20742 USA}

\author{S. Subhankar}
\affiliation{Joint Quantum Institute, NIST/University of Maryland, College Park, Maryland 20742 USA}

\author{Y. Wang}
\affiliation{Joint Quantum Institute, NIST/University of Maryland, College Park, Maryland 20742 USA}

\author{T-C Tsui}
\affiliation{Joint Quantum Institute, NIST/University of Maryland, College Park, Maryland 20742 USA}

\author{F.~Jendrzejewski} \affiliation{Kirchhoff Institut f\"ur Physik, Universit\"at Heidelberg, Im Neuenheimer Feld 227, 69120 Heidelberg, Germany}

\author{T.~Tiecke}
\email{Present address: Facebook, Inc. 1 Hacker Way, Menlo Park, CA, 94025}
\affiliation{Department of Physics, Harvard University, Cambridge, Massachusetts 02138, USA}

\author{G. Juzeli\=unas} \affiliation{Institute of Theoretical Physics and Astronomy, Vilnius University, Saul\.{e}tekio Avenue 3, LT-10257
Vilnius, Lithuania}

\author{L. Jiang} \affiliation{Department of Applied Physics, Yale University, New Haven, Connecticut 06520, USA}
\affiliation{Yale Quantum Institute, Yale University, New Haven, Connecticut 06511, USA}

\author{S. L. Rolston}
\affiliation{Joint Quantum Institute, NIST/University of Maryland, College Park, Maryland 20742 USA}

\author{J. V. Porto}
\affiliation{Joint Quantum Institute, NIST/University of Maryland, College Park, Maryland 20742 USA}

\author{A. V. Gorshkov}
\affiliation{Joint Quantum Institute, NIST/University of Maryland, College Park, Maryland 20742 USA}
\affiliation{Joint Center for Quantum Information and Computer Science, NIST/University of Maryland, College Park, Maryland 20742 USA}


\begin{abstract}
There has been a recent surge of interest and progress in creating \subwavelength free-space optical potentials for ultra-cold atoms.
A key open question is whether geometric potentials, which are repulsive and ubiquitous in the creation of \subwavelength free-space potentials, forbid the creation of narrow traps with long lifetimes. 
Here, we show that it is possible to create such traps.
We propose two schemes for realizing \subwavelength traps and demonstrate their superiority over existing proposals.
We analyze the lifetime of atoms in such traps and show that long-lived bound states are possible. 
This work opens a new frontier for the \subwavelength control and manipulation of ultracold matter, with applications in quantum chemistry and quantum simulation. 
\end{abstract}

\maketitle
Coherent manipulation of atoms using light is at the heart of cold-atom-based quantum technologies such as quantum information processing and quantum simulation~\cite{Gross2017,Lewenstein2012}. The most commonly used methods to trap atoms optically are based on the AC Stark shift induced in a two-level system by an off-resonant laser field, which provides a conservative potential that is proportional to  laser intensity.
The spatial resolution of such a trapping potential is diffraction-limited, unless operated near surfaces \cite{Gonzalez-Tudela2015,Mitsch2014,Thompson2013,Chang2009,Gullans2012,Romero-Isart2013}. 
In contrast, a three-level system with two coupling fields offers more flexibility and can generate a \subwavelength optical potential even in the  far-field:  although the intensity profiles of both laser beams involved are diffraction-limited, the internal structure of the state can change in space on length scales much shorter than the wavelength $\lambda$ of the lasers~\cite{Agarwal2006,Bajcsy2003,Dutton2001,Gorshkov2008a,Juzeliunas2007,Miles2013,Sahrai2005,Yavuz2007,Subhankar2018,Mcdonald2018}. 
Such \subwavelength internal-state structure can lead to \subwavelength potentials either by creating spatially varying sensitivity to a standard AC Stark shift~\cite{Yavuz2009} or by inducing a conservative \subwavelength geometric potential~\cite{Lacki2016,Jendrzejewski2016,Wang2018}.

Trapping atoms in the far field on the subwavelength scale
may allow for the realization of Hubbard-type models with increased tunneling and interaction energies \cite{Gullans2012,Romero-Isart2013,Gonzalez-Tudela2015,Yi2008,Lundblad2008}, which in turn would relax requirements on the temperature and coherence times in such experiments. 
Subwavelength traps can also be useful in atom-based approaches to quantum information processing~\cite{Perczel2017b,Grankin2018} and quantum materials engineering,  as well as for efficient loading into traps close to surfaces~\cite{Gonzalez-Tudela2015,Mitsch2014,Thompson2013,Chang2009,Gullans2012,Romero-Isart2013}. 
The use of dynamically adjustable \subwavelength tweezers~\cite{Barredo2016,Endres2016}, in which atoms can be brought together and apart, can also enable controlled ultracold quantum chemistry~\cite{Luhmann2015,Ospelkaus2010,Liu2018a}. 

To trap atoms on a subwavelength scale, the optical potential must provide a local minimum. The geometric scalar potential associated with laser-induced internal-state structure is always repulsive and increases in magnitude as its spatial extent is reduced. This repulsive contribution must be considered when engineering attractive \subwavelength optical potentials. A trap based on the combination of AC Stark shift and \subwavelength localization~\cite{Gorshkov2008a,Bajcsy2003,Dutton2001,Sahrai2005,Agarwal2006,Yavuz2007,Juzeliunas2007,Miles2013,
Kapale2013,Johnson1998,Thomas1989,Stokes1991,Schrader2004,Gardner1993,Zhang2006,Lee2007,Kapale2010,Kien1997,
Qamar2000,Paspalakis2001,Hell2007,
Li2008,Mompart2009,Sun2011,Proite2011,Qi2012,Viscor2012,Yavuz2012,Rubio2013} 
within a three-level system was proposed in Ref.~\cite{Yavuz2009}, but the geometric potentials arising from non-adiabatic corrections to the Born-Oppenheimer approximation~\cite{Lacki2016,Jendrzejewski2016} were not considered. In this \paper, we show that even with the repulsive
non-adiabatic corrections, attractive \subwavelength potentials are still possible. We also propose two alternative schemes for the generation of traps that offer significantly longer trapping times as compared to the approach of Ref.~\cite{Yavuz2009}. 
We analyze the performance of all three approaches and show that 8nm-wide traps offering 10ms trapping times are within reach.
Compared with near-field methods, our far-field approach not only avoids losses and decoherence mechanisms associated with proximity to surfaces, but also provides more flexibility in time-dependent control of the shape and position of the trapping potentials and, additionally, works not only in one and two but also in three dimensions.

\begin{figure}[h!] \includegraphics[width=0.95\columnwidth]{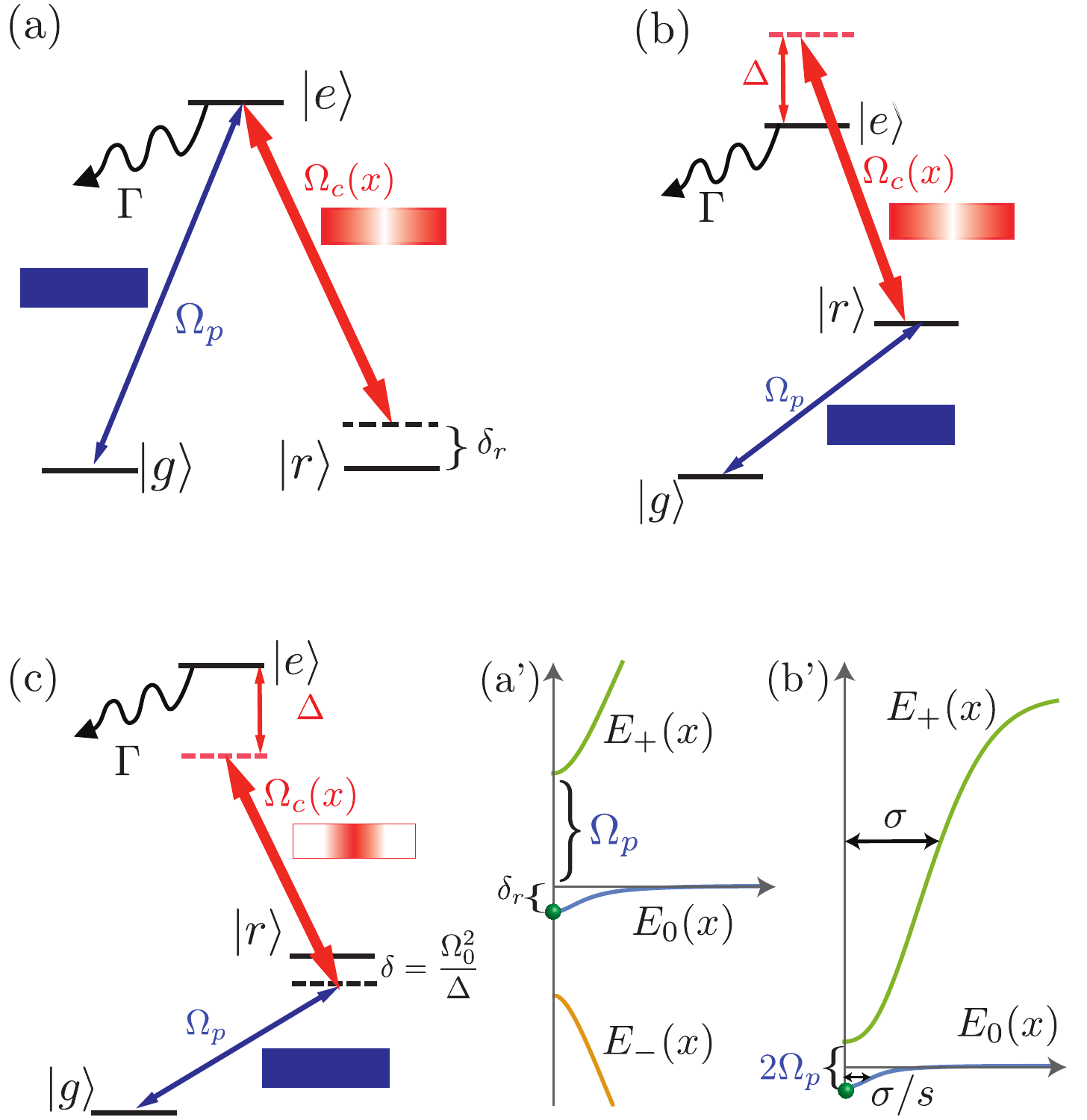}
 \caption{
(a) Level diagram for  the \EIT scheme, showing a spatially homogeneous field $\Omega_p$ [blue bar] and a spatially varying field $\Omega_c(x)=\Omega_0(1-e^{-x^2/\sigma^2})^{1/2}$ [red bar with gradient]. $\Gamma$ is the linewidth of the excited state.
 (b) Level diagram for the \Rf scheme. 
The intermediate state $\ket{r}$  is dressed by coupling it to the excited state $\ket{e}$  with a spatially dependent $\Omega_c (x) =\Omega_0(1-e^{-x^2/\sigma^2})^{1/2}$  and a large detuning $\Delta\gg\Omega_c$, which gives rise to a \lightshift $\Omega^2_c(x)/\Delta$ of state $\ket{r}$. 
The ground state $\ket{g}$ is coupled to state $\ket{r}$ with a spatially uniform $\Omega_p$ and detuning $\delta=0$. 
(c) Level diagram for the \Yavuz scheme~\cite{Yavuz2009}. 
The difference from (b) is that $\Omega_c (x) = \Omega_0e^{-x^2/(2 \sigma^2)}$ is maximal at $x = 0$ and that $|\Delta|\gg\Omega_c,\delta$ now indicates the amount of red detuning. 
Moreover, the detuning $\delta= \Omega^2_0/\Delta$ is chosen to exactly compensate for the \lightshift of $\ket{r}$ at $x = 0$.
(a',b') Sketches of the relevant eigenstates (atom depicted by a green ball is trapped in the blue potential): (a') for the (a) scheme;
(b') for the (b) scheme, which for  $x<w$ is equivalent to the (c) scheme. Although $E_{\pm}$ are diffraction limited, $E_0$ has \subwavelength shape characterized by width $w$, which can be expressed using the enhancement factor defined as $s=\sigma/w$.
}
\label{fig:scheme}
\end{figure}

\textit{Model.}---We start with a single-atom Hamiltonian \beqa
H=H_{\rs al}(x)+\frac{p^2}{2m},
\label{generalHam}
\eeqa
where $m$ is the mass, $p$ is the momentum, and $H_{\rs al}$ describes the atom-light interaction. 
We will consider three schemes shown in \figref{fig:scheme}: (a) electromagnetically induced transparency (\EIT), (b) \Rf, and (c) \Yavuz~\cite{Yavuz2009}.
For the \EIT scheme ($\hbar=1$), \beqa
H_{\rs al}=
\left(
\begin{array}{ccc}
\delta_r& 0 & \Omega _c(x) \\
0&0& \Omega_p  \\
 \Omega _c(x) & \Omega_p  & \Delta\\\end{array}
\right)\label{HaEIT}
\eeqa
in the basis of bare atomic states $\{\ket{r},\ket{g}, \ket{e}\}$, where $2\Omega_p$ and $2\Omega_c(x)$ are Rabi frequencies of a spatially homogeneous probe field and a spatially-varying control field, respectively.  
For the two AC-Stark schemes, in the limit of large single-photon detuning $|\Delta|\gg\Omega_c(x),\Omega_p,|\delta|$ [see Fig.\ \ref{fig:scheme}(b,c)], the intermediate state $\ket{e}$ can be adiabatically eliminated, resulting in an effective two-state Hamiltonian
\beqa
H_{\rs al}=
\left(
\begin{array}{cc}
\delta-\frac{\Omega_c^2(x)}{\Delta} & \Omega_p  \\
\Omega_p&  0
\end{array}
\right)\label{HaEliminated}
\eeqa
in the $\{\ket{r},\ket{g}\}$ basis. 

Within the Born-Oppenheimer approximation, we first diagonalize $H_{\rs al}$ which leads to  position-dependent eigenstates. Non-adiabatic corrections give rise to geometric scalar $\UNA$ and vector ${A}$ potentials, defined as $\UNA = R^\dagger\partial_x^2 R$ and $A = iR^\dagger\partial_xR$,
where $R$ is a unitary operator diagonalizing $H_{\rs al}$~\cite{Lacki2016,Jendrzejewski2016}. The resulting Hamiltonian is given by
$H'=R^\dagger HR=R^\dagger H_{\rs al}R+\UNA(x)+\frac{p^2}{2m}-\frac{A(x)p}{m}$.
Below, we focus on the potential $R^\dagger H_{\rs al}R+\UNA(x)$ experienced by three-level atoms under three different schemes.

\textit{\EIT scheme.}---In Refs.~\cite{Lacki2016,Jendrzejewski2016,Wang2018}, \subwavelength barriers were considered in the \EIT  configuration assuming two-photon resonance, i.e. $\delta_r=0$ in \figref{fig:scheme}(a). 
The approximate dark state $\ket{D}\propto \Omega_c(x)\ket{g}-\Omega_p\ket{r}$ then experiences only a repulsive geometric potential $\bra{D}U\ket{D}$.
On the other hand, in the presence of a finite detuning $\delta_r$ for  state $\ket{r}$,
the dark state $\ket{D}$ can acquire a negative energy shift $E_0(x)$ with an absolute value greater than the positive geometric potential. 
Moreover, we see that, as we move from large to small $x$, the state $\ket{D}$ changes its character from $\ket{g}$ to $\ket{r}$ at $x=w$ defined via $\Omega_c(w)=\Omega_p$.
Therefore, for $\Omega_0\gg \Omega_p$, we can engineer \subwavelength traps with width $w\ll\sigma$.	
However, at first glance, it is not obvious whether the additional contribution from the repulsive geometric potential would cancel the attractive potential. 
Moreover, the approximate dark state experiencing the trapping potential can have a significant admixture of state $\ket{e}$, leading to loss.
Below, we address these two issues.

In the following, for simplicity, we set $\Delta=0$ because, for a single trap  in the \EIT configuration, nearly all results (except the tunneling losses to the lower dressed-state $\ket{-}$) are $\Delta$-independent. 
  For $|\delta_r+U(x)|\ll\Omega_p$, the bright states $\ket{\pm}$ are well-separated from the dark state.
In this case, the ground state is composed of the dark state with a small admixture of bright states, so that the geometric potential and the energy shift $E_0$ can be calculated separately, see \figref{fig:scheme}(a'). 
Note that, for all schemes, we will take into account decay $\Gamma$ of state $\ket{e}$ perturbatively. 
We are interested in a spatially dependent~\footnote{
Such an intensity profile can be approximately implemented by using existing techniques such as intensity masks~\cite{Eckel2014}, Hermite-Gaussian laser modes, or holographic techniques~\cite{Gaunt2013}.} 
 control Rabi frequency
$\Omega_c (x) =\Omega_0(1-e^{-x^2/\sigma^2})^{1/2}$. 
For small $x$, $\Omega_c \approx \Om_0 x/\sigma \label{linear}$,
 so that the total effective potential $V_{\rs tot}= |\!\braket{D}{r}|^2 \delta_r+U_D$ is equal to 
\begin{equation}
V_{\rs tot}
=\frac{\delta_r}{1+x^2/w^2}+\frac{1}{2mw^2}\frac{1}{\left(1+x^2/w^2\right)^2},
\end{equation}
where we used $ U_D=\bra{D}U\ket{D} = \frac{1}{2 m} \left(\frac{\Omega_p \, \partial_x \Omega_c(x)}{\Omega_p^2+\Omega_c^2(x)}\right)^2$ 
and $w=\sigma\Omega_p/\Omega_0$. We see explicitly that the trapping potential has \subwavelength width $w$, which can be characterized by the enhancement factor $s=\sigma/w$, and that $U_D$ is always repulsive. 

To compare all three schemes, we start by considering traps that have a specific width $w$ and support a single bound state. Furthermore, we assume that our maximum Rabi frequency  $\Omega_c(x)$ is limited to $\Omega_0$. In that case, if we drop factors of order unity, our scheme supports a single bound state when the kinetic energy  $E_w=1\hbarSq/( 2mw^2)$ is equal the depth of the potential  $V_{\rs tot}$.

The leading source of loss comes from the admixture of the short-lived state $\ket{e}$. 
There are two processes leading to this admixture: (1) imperfect EIT due to $\delta_r \neq 0$ and (2)
 non-adiabatic off-diagonal corrections. Both processes admix $\ket{D}$ with $\ket{\pm}$, which in turn have significant overlap with $\ket{e}$. Within second-order perturbation theory, the loss rates from processes (1) and (2) are  $\Gamma_D^{(1)}
\sim 
\Gamma V_{\rs tot}^2/\Omega_p^2$ and
$\Gamma^{(2)}_D\sim \Gamma \frac{U_{D\pm}}{\Omega_p^2}\sim \Gamma \frac{E_w^2}{\Omega_p^2}$, respectively. Here $U_{D\pm}=\bra{D}U\ket{\pm}$
and we used the fact that, for a trap with a single bound state, the off-diagonal~\cite{Jendrzejewski2016}  terms of $\UNA$ are of the same order as $E_w$. Thus, up to factors of order unity, the total losses are 
$\Gamma_D\sim \Gamma_D^{(1)}+\Gamma_D^{(2)}\sim \Gamma E_w^2/\Omega_p^2$. 
We would like to note that we can modify the \EIT setup so that non-adiabatic corrections are further suppressed~\cite{supplementTraps} and the only (and unavoidable) losses come from imperfect EIT. The decay rate for the bound state can be expressed using $\ER$, $\Omega_0$, and $s$ as
$\Gamma_D\sim \Gamma s^6(\ER/\Omega_0)^2$, where $\ER \sim  1\hbarOne/(2m\sigma^2)$. An additional constraint on available widths $w$ comes from the fact that 
our perturbative analysis holds only for  $|V_{\rs tot}|$ and $E_w$ much smaller than the gap to the bright states $\ket{\pm}$, leading to $E_w\ll\Omega_p$, which is equivalent to $s^3\ll\Omega_0/\ER$.
Another source of losses is tunneling from the \subwavelength-trapped  state~\cite{Yi2008} to state $\ket{-}$, which, based on a Landau-Zener like estimate~\cite{supplementTraps}, is negligible for 
$s^3\ll\Omega_0/\ER$. The specific experimental parameters will be analyzed after the presentation of all three schemes.

\textit{\RfCapital  scheme.}---The second new schemes we propose is shown in \figref{fig:scheme}(b) and is described by  the Hamiltonian \eqref{HaEliminated} with $\delta=0$.
Here, the intermediate state $\ket{r}$  is dressed by coupling it to the excited state $\ket{e}$  with a spatially dependent Rabi frequency 
$\Omega_c (x) =\Omega_0(1-e^{-x^2/\sigma^2})^{1/2}$. 
Together with a large blue detuning $|\Delta|\gg\Omega_c(x)$, this leads to a \lightshift $\Omega^2_c(x)/\Delta$ of state $\ket{r}$. 
At large $x$, state $\ket{0}$ is equal to $\ket{g}$; whereas, at $x=0$, it is proportional to $\ket{g}-\ket{r}$.
The \lightshift $E_0$ describing the trapped state $\ket{0}$ is equal to
\beqa
E_0(x)=\Omega_p \left(
\frac{1}{2} \left(\frac { x} w\right)^2-\sqrt{1+\frac{1}{4} \left(\frac{x}{w}\right)^4}
\right),
\label{E0AC}
\eeqa
where the width $w$ equals $ \sigma/s$ with
  \beqa
  s =\sqrt{\frac{\Omega_0^2}{|\Delta|\Omega_p}}.\label{sForAC}
  \eeqa
   Intuitively, the width $w$ is equal to the distance at which the AC-stark shift is equal to the coupling $\Omega_p$.
   
For this scheme, non-adiabatic potential $\UNA$ is equal to  \beqa
U=
\left(
\begin{array}{cc}
 \alpha  & -\beta  \\
 \beta  & \alpha  \\
\end{array}
\right)
\label{eq:Udef}
\eeqa
with $\alpha=E_w\frac{4 w^2 x^2}{\left(4 w^4+x^4\right)^2}$ and $\beta=E_w\frac{6  x^4-8 w^4}{\left(4 w^4+x^4\right)^2}$. \begin{figure}[t]
 \includegraphics[width=0.48\columnwidth]{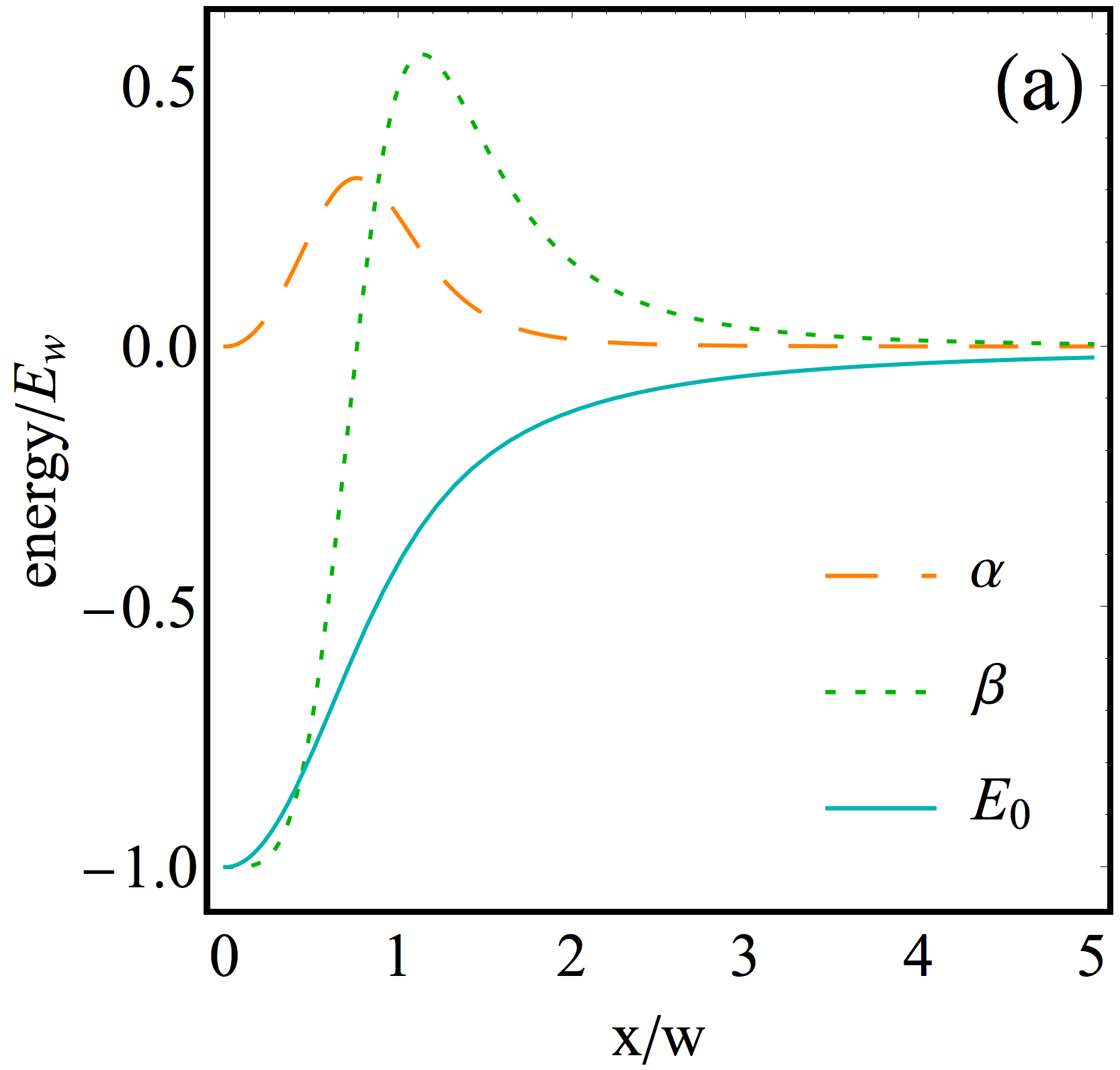}
  \hspace{.2em}
  \includegraphics[width=0.48\columnwidth]{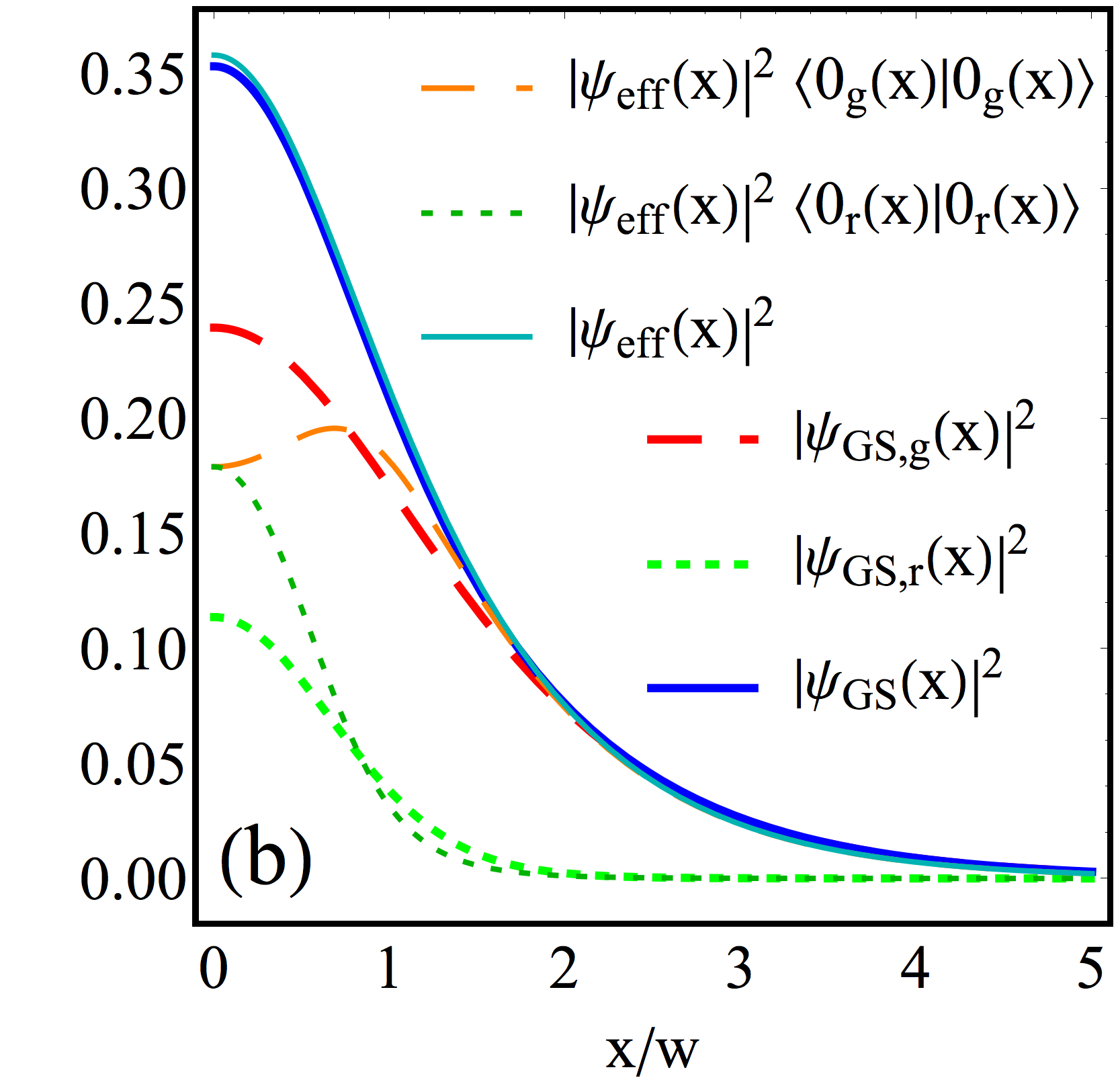}
  \caption{ 
Analysis of the \Rf scheme [Fig.\ \ref{fig:scheme}(b)]. 
(a) The \lightshift $E_0(x)$, as well as the diagonal and off-diagonal couplings coming from the non-adiabatic potential $U$ in Eq.~\eqref{eq:Udef} parametrized by $\alpha$ and $\beta$. (b) Properties of the ground state obtained from the effective Hamiltonian and from the full Hamiltonian  (see main text for details).
Figures are shown in units of $w$ and $E_w$ making them applicable to all $s \gg 1$.
}
\label{fig:Rf}
\end{figure}
 Note that the off-diagonal terms are significantly greater than the diagonal ones (i.e., $\alpha<|\beta|$), especially  for $x\lesssim w$, as shown in \figref{fig:Rf}(a). For $\Omega_p= E_w$, which leads  to a single bound state, we obtain  $\beta$ on the order of the energy $E_0(x)$. Note that our derivation works for arbitrary fractional probabilities 
$f_r=|{\psi_{r}(x)}/ {\psi_{}(x)  } |$, 
whereas the method in Ref.~\cite{Yavuz2009} works only for fractional probabilities $f_r\ll1$, where $\psi_r=\braket{r}{\psi}$ is the $r$-component of the ground-state wave function $\psi$. 

In order to analyze the impact of $\UNA$, we compare the ground-state of the effective Hamiltonian 
$H_{\rs eff}=E_0(x)-\frac{\partial_x^2}{2m}
$ without $\UNA$ with the exact solution of the full Hamiltonian given by Eqs.~\eqref{generalHam} and \eqref{HaEliminated}.
Even though  $|\bra{0}\UNA\ket{+}|\sim E_w\sim\Omega_p$  is large and on the order of  the energy difference $E_+-E_0 \sim \Omega_p$,  we see in  \figref{fig:Rf}(b) that the probability densities (and therefore the widths) of the ground states $\psi_{\rs eff}$ of $H_{\rs eff}$ and $\psi_{\rs GS}$ of the full Hamiltonian are nearly the same. 
However, from the comparison of components $\ket{g}$ and $\ket{r}$ of the ground state in Fig.~\ref{fig:Rf}(b), we see that the trapped atoms are not exactly in the eigenstate $\ket{0}$.
This partially explains why the non-adiabatic corrections do not influence the width of the ground state: the components of  the true ground state are smoother (spatial gradients are smaller) than than those of the ground state $\ket{0}$ of $H_{\rs al}$, which leads to weaker non-adiabatic corrections for the true ground state.
 In summary, even though the non-adiabatic potential $U$ can be on the order of $E_w$ for \subwavelength traps, the width of the ground state is only very weakly influenced by $U$.   
 
We now turn to the analysis of the trap lifetime. The leading contribution to losses comes from the admixture $P_e$ of the short-lived state $\ket{e}$. $P_e$ is determined by the characteristic coupling strength $\Omega_c(w)\approx \Omega_0/s$ within the trapped region and by the detuning $\Delta$ as $P_e \sim (\Omega_0/(\Delta s))^2 \sim  s^6(\ER/\Omega_0)^2$.
In principle, the condition $\Delta > \Omega_0$ might  give an upper limit on $s$, which, based on Eq.~\eqref{sForAC}, for $\Omega_p = E_w$, is $s^4< \Omega_0/E_\sigma$. However, this is not a constraint for any of the results considered in this Letter.

\textit{\YavuzCapital scheme.}---Finally, we analyze the third scheme, which was proposed in Ref.~\cite{Yavuz2009}. 
Our analysis, compared to the original one, takes into account non-adiabatic corrections and works for arbitrary fractional probabilities.
This scheme differs from the \Rf scheme in that: first, the control Rabi frequency is $\Omega_c(x) =\Omega_0e^{-x^2/(2 \sigma^2)}$ which, for small $x$, is $\approx \Omega_0(1-x^2/(2 \sigma^2))$; second,  the 
detuning $\delta= \Omega^2_0/\Delta$ is chosen to exactly compensate for the AC-Stark shift at the center of the trap~\footnote{Depending on the desired parameters of the trap, choosing a slightly larger detuning $\delta$ can sometimes slightly improve the scheme by achieving the optimal trade off between non-adiabaticity and scattering. However, the improvement is insignificant, so we chose to focus on 
$\delta=\Omega_0^2/\Delta$ to simplify the presentation.}; third, the detuning $\Delta$ now indicates the amount of red detuning.
The resulting {$E_0$} $w$, and $s$ are identical to those in the \Rf scheme, Eqs.~\eqref{E0AC} and \eqref{sForAC}.
We find that, for $x\lesssim w$, the non-adiabatic corrections have nearly exactly the same form as in the \Rf scheme and differ only in the sign of the off-diagonal terms:
$U=
\left(
\begin{array}{cc}
 \alpha  & \beta  \\
 -\beta  & \alpha  \\
\end{array}
\right).
$

To derive the lifetime of this trap, we can set $\Omega_c(x)$ to $\Omega_0$ within the trapped region,  which leads to
$P_e \sim (\Omega_0^2/\Delta^2)=(s^2\Omega_p/\Omega_0)^2 =s^8(\ER/\Omega_0)^2$.
This expression is identical to the one in the \EIT and \Rf schemes, except for the more favorable scaling with $s$ in our two schemes ($s^6$ vs.\ $s^8$), making them superior. 
The intuition behind the difference between the two schemes based on the AC-Stark shift is the following: in the \Yavuz scheme, the atoms are trapped in the region of maximal scattering from state $\ket{e}$, whereas, in our \Rf scheme, atoms are trapped in the region of minimal scattering from state  $\ket{e}$. 

\begin{figure}[t]
\includegraphics[width=0.85\columnwidth]{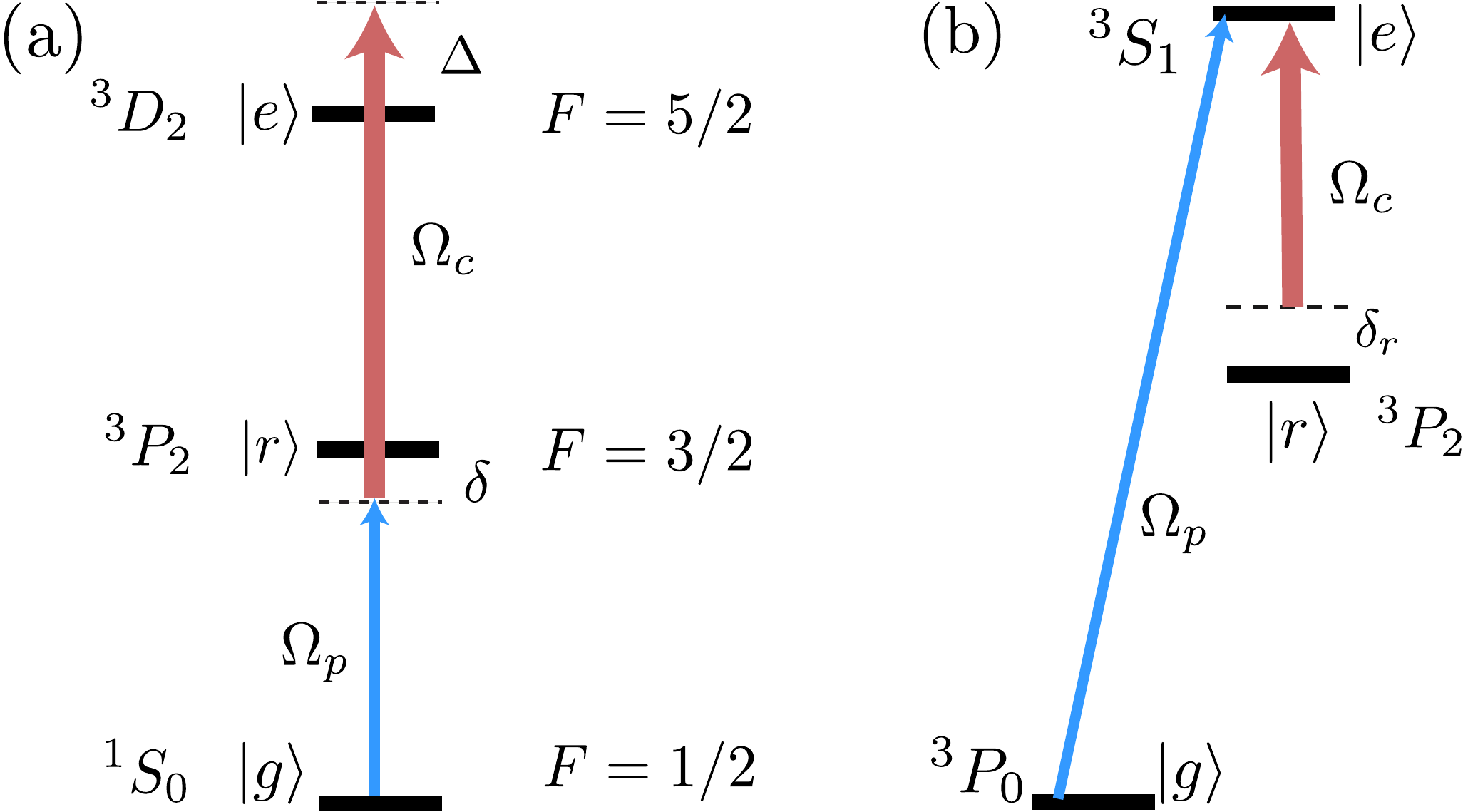}
\caption{
(a) 
Atomic levels for the two AC-Stark schemes. The decay rate from the metastable state $^3P_2$ is negligible ($\Gamma_{\rs 3P2}/2\pi=0.02$Hz). 
 (b)
Atomic levels for the \EIT scheme.
In all schemes, the main limitation comes from the admixture of levels outside the 3-level system. }
\label{fig:levels}
\end{figure}

\textit{Atomic levels.}---The level structure needed for the two AC-Stark schemes is most easily achieved with alkaline-earth atoms, in which $\ket{g}$, $\ket{r}$, and $\ket{e}$ are chosen to be the ground state $^1S_0$, the metastable state $^3P_2$, and the state ${}^3D_2$,  respectively~\footnote{The transition $^1S_0-\,^3P_2$ has greater matrix element~\cite{Porsev2004} than transition $^1S_0-\,^3P_0$, which makes it possible to work with greater $\Omega_p$.}, see~\figref{fig:levels}(a). 
The optical separation between the two long-lived states allows the decoupling of $\Omega_c$ from $\ket{g}$ to be a much better approximation~ \footnote{
The main limitation comes from  the off-resonant coupling of $\ket{g}$ by the $\Omega_c(x)$ field to  $^3P_1$, $^3D_2$, and	 $^1P_1$.} than what is possible in alkali atoms, where the size of $\Delta$ is limited by the fine structure splitting between the D1 and D2 lines~\cite{Yavuz2009}.

Turning now to the \EIT scheme, 
the subwavelength trap depths achievable with 
the atomic levels used for barriers in Ref.~\cite{Yang2018} are limited due to the off-resonant $\Omega_c$-induced coupling of $\ket{g}$ to $\ket{^3P_1, F=3/2}$~\footnote{More precisely, $\Omega_c$ couples $\ket{g}$=$\ket{^1S_0,F=1/2,m_F=-1/2}$ to $\ket{^3P_1,F=3/2,m_F=-3/2}$.}, which is detuned by $\Delta_{\rs hfs}/2\pi=5.94$GHz from  $\ket{^3P_1,F=1/2}$. 
This coupling gives rise to a position-dependent \lightshift of $\ket{g}$ and leads to an additional constraint $\Omega_c\ll\sqrt{\Delta_{\rs hfs}E_w}$ for trap realization. A solution [similar to the one used for the two AC-Stark schemes] is to protect our three-level system by an optical separation, as shown in~\figref{fig:levels}(b)
~\footnote{The main limitation comes from the coupling of  $\ket{g}$ to the $^3P_1$, $^3D_2$, and $^1P_1$  manifolds.}.

Note that the atomic level configurations in~\figref{fig:levels} do not rely on optical polarization selection rules.  Therefore, unlike the level configuration of Ref.~\cite{Wang2018}, such subwavelength traps can be extended into 3D.

\textit{Achievable trap parameters.}---We showed above that, for fixed $\Omega_0$, the two schemes proposed in this Letter provide superior performance to the \Yavuz scheme due to the $s^6$ vs.\ $s^8$ scaling of the losses.
We now discuss what widths of the trapping potentials are achievable when we include fundamental limitations imposed on the magnitude of $\Omega_0$.
We set the trapping time $T$ to be equal~\cite{Gonzalez-Tudela2015}
to $10$ms, and consider $^{171}$Yb. Depending on the scheme and on $\sigma$ [equal to $\lambda/2\pi$ for the lattice, and to $3\mu m$ for the tweezer; denoted by subscripts $_{\lambda}$ and $_{3\mu m}$, respectively], we find maximal $\Omega_0$ and $s$ such that the off-resonant position-dependent \lightshifts are less than $0.1 E_w$ and that $T\Gamma P_e\sim 1$:
\beqa
\begin{array}{c|c|cc|c|cc}
 \text{setup} & \frac{\Omega _{0,\lambda }}{2\text{$\pi $GHz}} & \frac{w_{\lambda }}{\text{nm}} & s_{\lambda } &
   \frac{\Omega _{0,3\text{$\mu $m}}}{2\text{$\pi $GHz}} & \frac{w_{3\text{$\mu $m}}}{\text{nm}} & s_{3\mu m}  \\
   \hline
 $\EITShort$ & 13 & 7.8 & 16. & 2.6 & 39. & 78. \\
 $\RfShort$ & 4.1 & 8.8 & 36. & 1.3 & 27. & 111. \\
 $\YavuzShort$ & 1.2 & 29. & 11. & 0.28 & 130. & 23.
\end{array}\nn
\eeqa
We see that the \EIT and the \Rf schemes allow for greater $\Omega_0$, which translates into narrower traps.
For comparison,  alkali-atom-based \EIT~\cite{Wang2018} and \Yavuz~\cite{Yavuz2009} schemes are limited to $\Omega_{0,\lambda}/2\pi=400\MHz$ (leading to $s=10$) and  $\Omega_{0,\lambda}/2\pi=150\MHz$  (leading to $s=1.7$), respectively. 

\textit{Applications.}---We now make a few remarks related to the applications pointed out in the introduction.
Note that, if one’s goal is simply to use the expansion of a control field $\Omega_c(x)$ around its nodes to create traps with tight bound states with minimal scattering, then our \EIT scheme has no advantages over a simple two-level blue-detuned trap. Indeed, in our case, up to an additive constant, the potential near a  node is given by $V(x) \approx\delta_r
 \Omega_c(x)^2/\Omega_p^2$, while the population of the excited state is given by $P_e(x) \approx \delta_r^2\Omega_c^2(x)/\Omega_p^4$. On the other hand, if one uses the same field $\Omega_c(x)$ to create a simple two-level blue-detuned trap (with detuning $\Delta$), one obtains $V(x) \approx \Omega_c^2(x)/\Delta$ and $P_e(x) \approx \Omega_c^2(x)/\Delta^2$. In other words, our scheme is identical to the two-level scheme provided one replaces $\Delta$ with $\Omega_p^2/\delta_r$. 

However, our goal is not only to create a tight bound state in a trap of \subwavelength width $w$ but also to make the trapping potential nearly constant for $|x| > w$ so that we can make and possibly independently move several traps, or a full lattice of traps, with \subwavelength separations. 
In that  case, a simple two-level scheme will not work. Instead, one has to use one of the \subwavelength schemes we discuss in this Letter.

In combination with stroboscopic techniques \cite{Nascimbene2015} or multi-level atomic schemes~\cite{Lacki2016}, our traps  can lead to the creation of lattices with \subwavelength periods, giving rise to large energy scales in Hubbard models~\cite{Gross2017,Cazalilla2009,Gorshkov2010,Daley2011,Cazalilla2014} and in dipolar atomic~\cite{Baier2016,Ferrier-Barbut2016,DePaz2013} and molecular~\cite{Micheli2006,Baranov2008,Pupillo2009,Carr2009a,Gorshkov2011c,Gorshkov2011a,Yan2013,Covey2015,Moses2015}
systems, with applications to quantum simulation and quantum computing.
Movable subwavelength traps with subwavelength separation may also find applications in ultracold chemistry~\cite{Luhmann2015,Ospelkaus2010,Liu2018a}.

\section*{Acknowledgments}
We are particularly grateful to Misha Lukin and Peter Zoller for stimulating discussions. We also thank Mikhail Baranov, Tommaso Calarco, Gretchen Campbell, Steve Eckel, Mateusz Lacki, Mingwu Lu, Jeff Thompson for helpful discussions.
Y.W., S.S., T-C. T., J.V.P., and S.L.R. acknowledge support by NSF PFC at JQI and ONR grant N000141712411. 
P.B. and A.V.G. acknowledge support by NSF PFC at JQI, AFOSR, ARL CDQI, ARO, ARO MURI, and NSF Ideas Lab. 
L.J. acknowledges support by ARL CDQI, ARO MURI, Sloan Foundation, and Packard Foundation.
F. J. acknowledges support by the DFG Collaborative Research Center ‘SFB 1225 (ISOQUANT)’, the DFG (Project-ID 377616843), the Excellence Initiative of the German federal government and the state governments—funding line Institutional Strategy (Zukunftskonzept): DFG project number ZUK49/Ü.

\bibliography{library}
\clearpage

\clearpage
\newpage
\section{Supplemental material}
\input{Supplement.tex}

\newpage 

\end{document}

%% file: Supplement.tex
%

\beginsupplement
\setcounter{equation}{0}
\renewcommand{\theequation}{S\arabic{equation}}

In Sec.~\ref{app:EIT}, we discuss how to modify the \EIT scheme to suppress non-adiabatic corrections.
In Sec.~\ref{app:LZ}, we estimate losses to lower dressed states.

\subsection{Modified \EIT scheme \label{app:EIT}}
Here, we show how to suppress non-adiabatic corrections in the \EIT scheme. 
The idea is that
$\Omega_c(0)$ does not necessarily have to go to zero and that the gradient of $\Omega_c(x)$ around $x=w$ can be smaller than for linear $\Omega_c(x)\sim x \Omega_0/\sigma$. Non-adiabatic corrections can then be suppressed by using the following control field~\cite{Yang2018}:
$\Omega_c(x) = \Omega_0(1+\nu-\cos(k x))$, which does not go to zero as deeply and as sharply as the linear $\Omega_c(x)$.

Expanding $\Omega_c(x)$ around a minimum for $\nu>0$, we find
\begin{equation}
\Om_c(x)=\Om_p(\eta+ (x/w)^2),
\label{OmcYang}
\end{equation} 
with $\eta=\nu\,\Omega_0/\Omega_p$ and $w=\frac 1 k\sqrt{2\Omega_p/\Omega_0} $, and
which gives rise to \beqa
V_{\rs tot}&=& \frac{\delta _r}{\left(\eta +(x/w)^2\right)^2+1}\! + \! \frac{4E_w (x/w)^2}{\left((\eta +(x/w)^2)^2+1\right)^2},\nn
\eeqa
whose depth can be tuned to accommodate one or more bound states.
By operating at $\eta>0$, we can use appropriate $|\delta_r|\sim E_w$ to engineer trapping potentials with negligible non-adiabatic potential $U$.
Therefore, when it comes to losses, this modified EIT scheme allows us to gain up to a factor of $\sim 2$.

\subsection{Landau-Zener estimates of losses to lower dressed states \label{app:LZ}}

Another source of losses is tunneling from the single bound state we consider to state $\ket{-}$. 
Note that, due to the conservation of energy, atoms in $\ket{-}$ will have large kinetic energy.
Following~\cite{Yi2008}, the loss rate   $\Gamma_{\rs LZ}$ can be estimated using a Landau-Zener like argument, which, in our setup, leads to
\beqa
\Gamma_{\rs LZ}\sim E_w e^{-\nu \Delta_{0-}/E_w},
\eeqa
where $\nu$ is a factor of order unity, and $\Delta_{0-}$ is the energy difference between two dressed states involved in the tunneling. 

In the EIT scheme, we have $\Delta_{0-}\sim |E_-(0)|\sim\Omega_p$ because the tunneling occurs  around $x\sim 0$ where the gap between $E_D$ and $E_-$ is smallest and where the atoms are trapped. This leads to the condition $1\ll \Omega_p/E_w=\Omega_c/(\ER s^3)$.
Note that we obtained the same condition from the requirement $E_w\ll \Omega_p$, which enabled us to treat non-adiabatic potentials and \lightshifts separately and perturbatively.  
We can further suppress tunneling losses by working at $\Delta\neq 0$.

In the \Rf scheme, $\Delta_{0-}\sim |E_-|\sim |\Delta|$, so this tunneling loss rate is strongly suppressed as $\exp[-|\Delta|/E_w]$.

In the \Yavuz scheme, there is no state below the state of interest and therefore no tunneling.